\documentclass[12pt,nofootinbib,superscriptaddress]{revtex4}

\usepackage{amssymb,color,paralist,amsmath,epsfig}
\usepackage{graphicx}
\usepackage{comment}
\usepackage{amsfonts}

\usepackage{relsize}

\newcommand{\cG}{{\cal G}}
\newcommand{\cF}{{\cal F}}
\newcommand{\cma}{{\theta_{\mathrm{cm}}}}

\newcommand{\hp}{{\frac{1}{2}}}
\newcommand{\hm}{{-\frac{1}{2}}}

\newcommand{\beq}{\begin{equation}}
\newcommand{\eeq}{\end{equation}}

\newcommand{\ber}{\begin{eqnarray}} 
\newcommand{\eer}{\end{eqnarray}}

\begin{document}

\title{ Two-photon exchange corrections in elastic muon-proton scattering }

\author{O. Tomalak}
\affiliation{Institut f\"ur Kernphysik, Johannes Gutenberg Universit\"at, Mainz, Germany}
\affiliation{PRISMA Cluster of Excellence, Johannes Gutenberg-Universit\"at,  Mainz, Germany}
\affiliation{Department of Physics, Taras Shevchenko National University of Kyiv, Ukraine}
\author{M. Vanderhaeghen}
\affiliation{Institut f\"ur Kernphysik, Johannes Gutenberg Universit\"at, Mainz, Germany}
\affiliation{PRISMA Cluster of Excellence, Johannes Gutenberg-Universit\"at,  Mainz, Germany}

\date{\today}

\begin{abstract}
We extend the general formalism of two-photon exchange to elastic lepton-nucleon scattering by accounting for all lepton mass terms. We then perform a numerical estimate of the muon-proton scattering at low momentum transfer in view of the future MUSE experiment. For this purpose, we estimate the two-photon exchange corrections to muon-proton scattering observables by considering the contribution of the proton intermediate state, which is expected to dominate at very low momentum transfers. We find that the two-photon exchange effect to the unpolarized muon-proton scattering cross section in the MUSE kinematical region is of the order of 0.5$\% $.
\end{abstract}

\maketitle

\section{Introduction}
\label{sec1}

The proton charge radius measurements from the hydrogen spectroscopy \cite{Mohr:2008fa} are in good agreement with the measurements from the unpolarized elastic electron-proton scattering \cite{Bernauer:2010wm}. In contrast, the recent measurements of the proton charge radius in the muonic hydrogen \cite{Pohl:2010zza, Antognini:1900ns} are in strong contradiction with the electronic results. This "proton charge radius puzzle" has not been solved yet.

One of the possible directions to understand the discrepancy is to verify the lepton universality by comparing measurements of the proton electromagnetic form factors in the unpolarized elastic muon-proton scattering with their counterparts using electron-proton scattering. The new muon-proton scattering experiment (MUSE) was proposed for these studies \cite{Gilman:2013eiv}. The precise determination of the proton charge radius requires an account of the two-photon exchange (TPE) corrections to the unpolarized elastic scattering. These corrections are expected to be of the order of $ 1 \% $ of the cross section.

An estimate of lepton mass effects in the elastic piece of the TPE corrections in lepton-nucleon scattering in the momentum transfer range $ 1\mathrm{-}2 ~\mathrm{GeV}^2 $ \cite{Chen:2013udl} showed a small difference between the corrections for the muon- and electron-proton elastic scattering.

For the momentum transfer range of the MUSE experiment $ Q^2 \lesssim 0.1 ~\rm{GeV}^2 $ we expect the main contribution to TPE corrections from the elastic, i.e. nucleon, intermediate state. In this work we develop the general formalism for elastic muon-proton scattering including TPE and estimate this contribution in a model with a proton intermediate state.

We introduce the general formalism of elastic muon-proton scattering in Sec. \ref{sec2}.  The evaluation of the two-photon box graph with the assumption of an on-shell virtual photon-proton-proton vertex is described in Sec. \ref{sec3}. We present results of our calculations for the MUSE kinematic region in Sec. \ref{sec4} and conclusions with outlook in Sec. \ref{sec5}.

\section{Elastic muon-proton scattering}
\label{sec2}

\begin{figure}[h]
\begin{center}
\includegraphics[width=.4\textwidth]{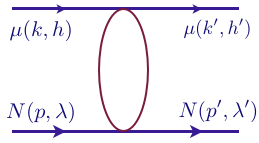}
\end{center}
\caption{Elastic muon-proton scattering.}
\label{mup_elastic}
\end{figure}

The kinematics of the elastic muon-proton scattering $ \mu( k , h ) + p( p, \lambda ) \to \mu( k', h') + p(p', \lambda') $, with $ h (h') $ and $ \lambda (\lambda') $ denoting the helicities of incoming (outgoing) muons and protons, respectively, see Fig. \ref{mup_elastic}, can be completely described by 2 Mandelstam variables, e.g., $ Q^2 = - (k-k')^2 $,  the squared momentum transfer, and $ s = ( p + k )^2 $, the squared energy in the muon-proton center-of-mass ( c. m. ) reference frame.

In the c. m. reference frame with muon scattering angle $ \cma $ the momentum transfer is given by
\ber
Q^2 = - ( k - k' )^2 = 2 | \vec{k} |^2 ( 1 - cos (\theta_{cm}) ) = \frac{ \Sigma }{2 s} ( 1 - cos (\theta_{cm}) )  ,
\eer
with $  \Sigma \equiv (s-(m+M)^2)(s-(m-M)^2) $, and $ m (M) $ denotes the muon (proton) mass respectively.  

In the laboratory frame with $ p = (M, 0), ~ k = (E,\vec{k}), ~ k' = (E',\vec{k'}), ~ p' = (E_p',\vec{k}-\vec{k}')$ the momentum transfer $  Q^2 = 2 M ( E - E') $ and  $ s = M^2 + m^2 + 2 M E $. The laboratory frame scattering angle $ \theta_{lab} $ and the momentum transfer are given by 
\ber
\cos \theta_{lab} & = & \frac{ E E' - m^2 - M (E - E')}{\sqrt{(E^2-m^2)(E'^2-m^2)}} ,  \\
Q^2 & = & 2 M \frac{( E^2 - m^2 )(  M + E \sin^2 \theta_{lab} - \sqrt{ M^2 - m^2 \sin^2 \theta_{lab} } \cos \theta_{lab})}{ ( E + M )^2 - ( E^2-m^2 ) \cos^2 \theta_{lab} } .
\eer

For the MUSE muon beam momenta $ k = 0.115, ~ k = 0.153, ~\mathrm{and} ~ k = 0.210 ~\mathrm{GeV} $ \cite{Gilman:2013eiv} the kinematically allowed momentum transfer is $ 0 < Q^2 < 4 M^2 ( E^2 - m^2 )/(m^2 + M ( 2 E + M ) ) $, or $ 0 < Q^2 < 0.039$, $ 0 < Q^2 < 0.066$ and $ 0 < Q^2 < 0.116 ~\mathrm{GeV}^2 $ respectively. For the scattering angles of the experiment $ 20^0 < \theta_{lab} < 100^0 $ the momentum transfer varies in the region $ 0.0016 - 0.026  $, $ 0.0028 - 0.045$ and $ 0.0052 - 0.080 ~\mathrm{GeV}^2 $ respectively. In the case of electron scattering with the same momenta and experimental scattering angles the momentum transfer varies in the region $ 0.0016 - 0.027 $, $ 0.0028 - 0.046  $ and $ 0.0052 - 0.082 ~\mathrm{GeV}^2 $.

It is convenient to introduce the averaged momentum variables $ P = (p+p')/2, ~~~ K = (k+k')/2 $, the $u$-channel squared energy $ u = ( k - p' )^2 $ and the crossing symmetric variable $ \nu = (s-u)/4 $ which changes sign with $ s\leftrightarrow u $ channel crossing. The crossing symmetric variable can be expressed in terms of the laboratory frame variables as $ \nu = M ( E + E' ) / 2$. Instead of the Mandelstam invariant $ s $ or the crossing symmetric variable $ \nu $, it can be convenient in experiment to use the virtual photon polarization parameter $ \varepsilon $, which varies between $ \varepsilon_0 = 2m^2/Q^2 $ and $1 $ for the momentum transfer $ Q^2 > 2 m^2 $ and between $ 1 $ and $ \varepsilon_0 $ for the momentum transfer $ Q^2 < 2 m^2 $. For the massless case, $ \varepsilon $ has the physical interpretation of the degree of the longitudinal polarization in the case of the one-photon exchange. The high energy limit corresponds to $ \varepsilon = 1 $. In terms of $ Q^2 $ and $ \nu $ the photon polarisation parameter is defined as
\ber
\varepsilon = \frac{16 \nu^2 - Q^2 ( Q^2 + 4M^2 )}{16 \nu^2 - Q^2 ( Q^2 + 4M^2 ) + 2 ( Q^2 + 4M^2 )( Q^2 - 2 m^2)} .
\eer

The value of the critical momentum transfer $ Q^2 = 2 m^2 $, corresponding to $ \varepsilon = 1 $ for all possible beam momenta, is given by $ Q^2 \simeq 0.022 \mathrm{GeV}^2 $. This value is inside the MUSE kinematic region for all three nominal beam momenta.

To describe lepton-nucleon scattering, there are 16 helicity amplitudes $ T_{h' \lambda', h \lambda} $ with arbitrary $h,h',\lambda,\lambda'$ = $\pm 1/2$ in Fig. \ref{mup_elastic}. The discrete symmetries of QCD and QED (parity and time-reversal invariance) leave just six independent amplitudes: $ T_1 = T_{\hp \hp, \hp \hp}, ~T_2 = T_{\hp \hm, \hp \hp},  ~T_3 = T_{\hp \hm, \hp \hm} , ~T_4 = T_{\hm \hp, \hp \hp}, ~T_5 = T_{\hm \hm, \hp \hp},  ~T_6 = T_{\hm \hp, \hp \hm} $. 

The helicity amplitudes for the $ l N $ elastic scattering can be expressed by the sum of six different tensor structures and generalized form factors (FFs). It is common to divide the helicity amplitudes into a part without  lepton helicity-flip which survives in the lepton massless limit $ T^{nonflip} $, and the part with lepton helicity-flip $ T^{flip} $ which is proportional to the mass of the lepton \cite{Gorchtein:2004ac} (where the T matrix is defined as $ S = 1 + i T $):
\ber \label{str_ampl} 
T^{nonflip} & = & \frac{e^2}{Q^2} \bar{u}(k',h') \gamma_\mu u(k,h) \cdot \bar{u}(p',\lambda') \left(\cG_M  \gamma^\mu - \cF_2  \frac{P^{\mu}}{M} + \cF_3  \frac{\gamma . K P^{\mu}}{M^2} \right) u(p,\lambda) , \label{str_ampl1} \\
 T^{flip} & = &\frac{e^2}{Q^2} \frac{m}{M} \bar{u}(k',h') u(k,h) \cdot \bar{u}(p',\lambda')\left( \cF_4  + \cF_5  \frac{\gamma . K}{M}\right) u(p,\lambda)  \nonumber \\
& + & \frac{e^2}{Q^2} \frac{m}{M} \cF_6  \bar{u}(k',h') \gamma_5 u(k,h) \cdot \bar{u}(p',\lambda') \gamma_5 u(p,\lambda) . \label{str_ampl2}
\eer

The helicity amplitudes can be expressed in terms of the generalized FFs and vice versa, as given in Appendix A.

In the one-photon exchange approximation the two surviving helicity amplitudes for $ \mu p $ elastic scattering can be expressed in terms of the Dirac $ F_1 $ and Pauli $ F_2 $ FFs
\beq \label{OPE_amplitude} 
T  =  \frac{e^2}{Q^2} \bar{u}(k',h') \gamma_\mu u(k,h) \cdot \bar{u}(p',\lambda') \left( \gamma^\mu F_1(Q^2) + \frac{i \sigma^{\mu \nu} q_\nu}{2 M} F_2(Q^2)  \right) u(p,\lambda).
\eeq
It is customary in experimental analyses to work with Sachs magnetic and electric FFs
\ber  \label{Sachs_ffs}
 G_M  =  F_1 + F_2 , ~~~~~~~
 G_E  = F_1 - \tau F_2,
\eer 
with $ \tau = Q^2 / (4 M^2) $. In the one-photon exchange approximation, the structure amplitudes defined in Eqs. (\ref{str_ampl1}), (\ref{str_ampl2}), can be expressed in terms of the one-photon exchange FFs $ \cG_M = G_M (Q^2), ~\cF_2 = F_2 (Q^2), ~\cF_3 = \cF_4 = \cF_5 = \cF_6 = 0  $. The exchange of more than one photon gives corrections of order $ O(\alpha) $, with $ \alpha = e^2/(4 \pi) \simeq 1/137 $, to all these amplitudes.

The TPE correction to the unpolarized elastic muon-proton cross-section is given by the interference between the one-photon exchange amplitude and the sum of box and crossed-box graphs with two photons. The correction can be defined through the difference between the cross section with accounts of exchange of two photons and the cross section in the $1 \gamma $-exchange approximation $ \sigma_{1 \gamma} $ by 
\ber
 \sigma = \sigma_{1 \gamma} \left( 1 + \delta_{2 \gamma} \right).
\eer
It can be expressed in terms of the TPE structure amplitudes as
\ber \label{delta}
\delta_{2\gamma}  & = & \frac{2}{ G_M^2 + \frac{\varepsilon}{\tau} G_E^2} \left\{ G_M \Re \cG_1 + \frac{\varepsilon}{\tau} G_E \Re \cG_2 + \frac{ 1 - \varepsilon }{ 1 - \varepsilon_0 } \left( \frac{\varepsilon_0}{\tau}  G_E \Re \cG_4 - G_M \Re \cG_3 \right) \right\} ,
\eer
where we defined for convenience the following amplitudes
\ber
 \cG_1 & = & \cG_M + \frac{\nu}{M^2} \cF_3 + \frac{m^2}{M^2} \cF_5,  \\
 \cG_2 & = & \cG_M - ( 1 + \tau ) \cF_2 + \frac{\nu}{M^2} \cF_3,  \\
 \cG_3 & = & \frac{m^2}{M^2} \cF_5 + \frac{\nu}{M^2} \cF_3,   \\
 \cG_4 & = &  \frac{\nu}{M^2}  \cF_4 + \frac{\nu^2}{M^4 (1+\tau)} \cF_5 . 
\eer
For the terms proportional to $ \cG_1 $, $ \cG_3 $, and $ \cG_4$ in Eq. (\ref{delta}), the contribution to $ \delta_{2 \gamma} $ starts from $ 0 $ when $ Q^2 $ vanishes. In this limit, the amplitude proportional to $ \cG_2$ dominates, and reduces in the massless lepton limit to the Feshbach correction \cite{McKinley:1948zz}. Note that the amplitude $ \cG_4 $ appears in the expression for the beam normal spin asymmetry up to the factor $  \frac{\nu}{M^2} $ \cite{Gorchtein:2004ac}. The contribution to $ \delta_{2 \gamma } $ which is linear in the amplitude $ \cF_6 $ vanishes, as well as its contribution to the beam normal spin asymmetry \cite{Gorchtein:2004ac}. The amplitude $ \cF_6 $ only  shows up in double polarisation observables.

\section{Box diagram model calculations}
\label{sec3}

\begin{figure}[htp]
\begin{center}
\includegraphics[width=.70\textwidth]{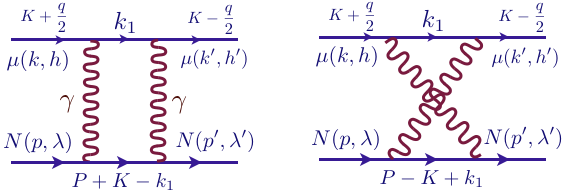}
\end{center}
\caption{Direct and crossed TPE diagrams.}
\label{model_graph}
\end{figure}

In this section, we will use a model to estimate the TPE effect to elastic muon-proton scattering at low momentum transfer. For such kinematics, we expect the dominant contribution to be given by the TPE direct box and crossed box diagram with proton intermediate state, as shown in Fig. \ref{model_graph}.

The helicity amplitude contribution from the direct and crossed TPE graphs (see Fig. \ref{model_graph}) can be expressed  as \cite{Blunden:2003sp}
\ber
\label{helamp}
 T_{direct}  = & &  - e^4 \mathop{\mathlarger{\int}} i \frac{  \mathrm{d}^4 k_1}{( 2 \pi )^4} \bar{u}(k',h') \gamma^\mu (\hat{k_1}+m) \gamma^\nu u (k,h) \bar{N}(p',\lambda') \Gamma_\mu (\hat{P} + \hat{K} - \hat{k}_1 + M) \Gamma_\nu N (p,\lambda) \nonumber \\
& & \frac{1}{(k_1 - P - K )^2 - M^2} \frac{1}{k_1^2 - m^2} \frac{1}{(k_1 - K - \frac{q}{2} )^2 - \mu^2 } \frac{1}{(k_1 - K + \frac{q}{2} )^2 - \mu^2 },  \\
 T_{crossed}  = & &- e^4 \mathop{\mathlarger{\int}} i \frac{ \mathrm{d}^4 k_1}{( 2 \pi )^4} \bar{u}(k',h') \gamma^\mu (\hat{k_1}+m) \gamma^\nu u (k,h) \bar{N} (p',\lambda') \Gamma_\nu (\hat{P} - \hat{K} + \hat{k}_1 + M) \Gamma_\mu N (p,\lambda) \nonumber \\
& & \frac{1}{(k_1 + P - K )^2 - M^2} \frac{1}{k_1^2 - m^2} \frac{1}{(k_1 - K - \frac{q}{2} )^2 - \mu^2 } \frac{1}{(k_1 - K + \frac{q}{2} )^2 - \mu^2 },
\eer
with the virtual photon-proton-proton vertex $ \Gamma^\mu $ and the infinitesimal photon mass $ \mu $, which regulates the IR divergencies. The structure amplitudes entering Eq. (\ref{delta}) can be expressed as combination of helicity amplitudes with the help of Eq. (\ref{stramp}).

The box diagram calculation was performed with the assumption of an on-shell form of the virtual photon-proton-proton vertex
\ber
\label{OPE2}
 \Gamma^\mu (Q^2) = \gamma^\mu F_1(Q^2) + \frac{i \sigma^{\mu \nu} q_\nu}{2 M} F_2(Q^2) ,
\eer
and by using (for simplicity) the dipole form for the proton electromagnetic FFs
\ber  \label{dipole_model}
 G_M & = & F_1 + F_2  = \frac{\kappa+1}{( 1 + Q^2/\Lambda^2 )^2}, \nonumber \\
 G_E & = &F_1 - \tau F_2  = \frac{1}{( 1 + Q^2/\Lambda^2 )^2} ,
\eer 
with $ \kappa = 1.793 $ and $ \Lambda^2 = 0.71 ~ \mathrm{GeV}^2 $.

\begin{figure}[htp]
\begin{center}
\includegraphics[width=.63\textwidth]{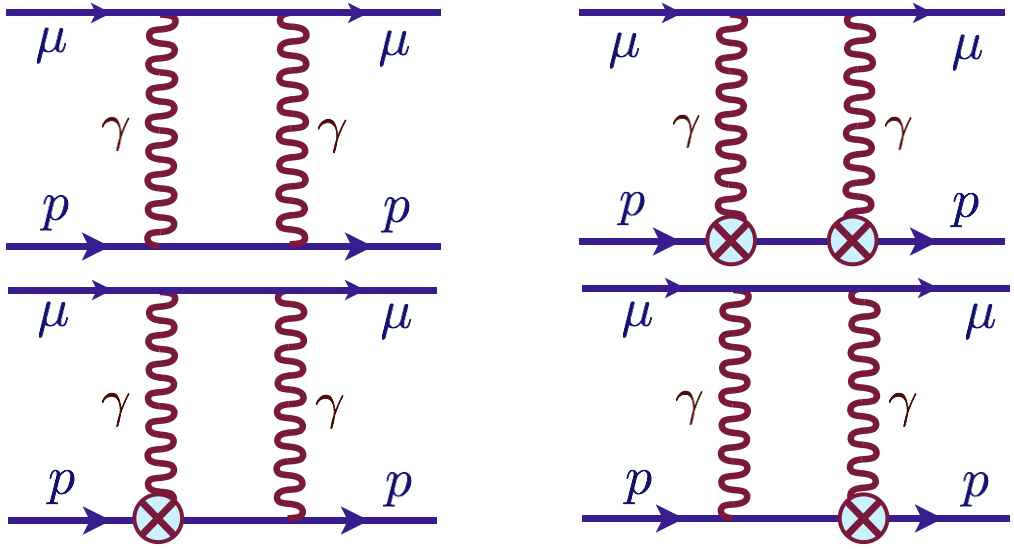}
\end{center}
\caption{F1F1 (upper left panel), F2F2 (upper right panel), and F1F2 (lower panels) structure of photon-proton-proton vertices. The vertex with (without) the cross denotes the contribution proportional to $ F_2 $ ($ F_1 $) form factor.}
\label{vertices}
\end{figure}

Due to the photon momentum in the numerator of the term multiplying the form factor $ F_2 $, the high energy behaviour of the amplitudes can be different depending on whether $ F_1 $ or $ F_2 $ enters the vertex. We denote the contribution with two electric coupling vertices by F1F1, two magnetic couplings by F2F2 and two contributions from the mixed case by F1F2 (see Fig. \ref{vertices}). The inclusion of FFs in the dipole form leads to a UV finite results for the structure amplitudes.

We used LOOPTOOLS \cite{Hahn:2000jm, vanOldenborgh:1989wn}  to evaluate the four-point integrals and derivatives from them, and to provide a numerical evaluation of the structure amplitudes. The calculation was done with the subtraction of the IR divergent term according to the Maximon and Tjon prescription \cite{Maximon:2000hm}. TPE amplitude $ \cG_M $ in the case of scattering of two point charges (i.e., F1F1 contribution with $ F_1 ( Q^2 ) = 1 $) has the IR divergent term 
\ber
 \cG^{IR,0}_M & = &    \frac{s - M^2 - m^2}{\sqrt{\Sigma}}   \left(\ln \left(\frac{\sqrt{\Sigma}-s+(m+M)^2}{\sqrt{\Sigma}+s-(m+M)^2}\right)+i \pi \right) \frac{\alpha}{\pi}  \ln \left(- \frac{t}{\mu^2}\right) \nonumber \\
 & - & \frac{u - M^2 - m^2}{\sqrt{\Sigma_u}}  \ln \left(\frac{\sqrt{\Sigma_u}-u+(m+M)^2}{ - \sqrt{\Sigma_u} -u + (m+M)^2}\right) \frac{\alpha}{\pi}  \ln \left(- \frac{t}{\mu^2}\right), 
\eer
with $  \Sigma_u \equiv (u-(m+M)^2)(u-(m-M)^2) $. The IR divergent contribution to $ \cG_M $ is given by $ F_1(Q^2) \cG^{IR,0}_M  $ for the F1F1 vertex structure. The IR divergent contribution to $ \cG_M $ and $ \cF_2 $ is given by $  F_2(Q^2) \cG^{IR,0}_M $ for the F1F2 vertex structure. The other amplitudes are IR finite in case of the F1F1 and F1F2 vertex structures. The F2F2 vertex structure is IR finite.

We also checked explicitly that the imaginary parts of the structure amplitudes evaluated through the box diagram calculation are in agreement with results of the calculation based on unitarity relations.

\section{Results and discussion}
\label{sec4}

\begin{figure}[htp]
\begin{center}
\includegraphics[width=.95\textwidth]{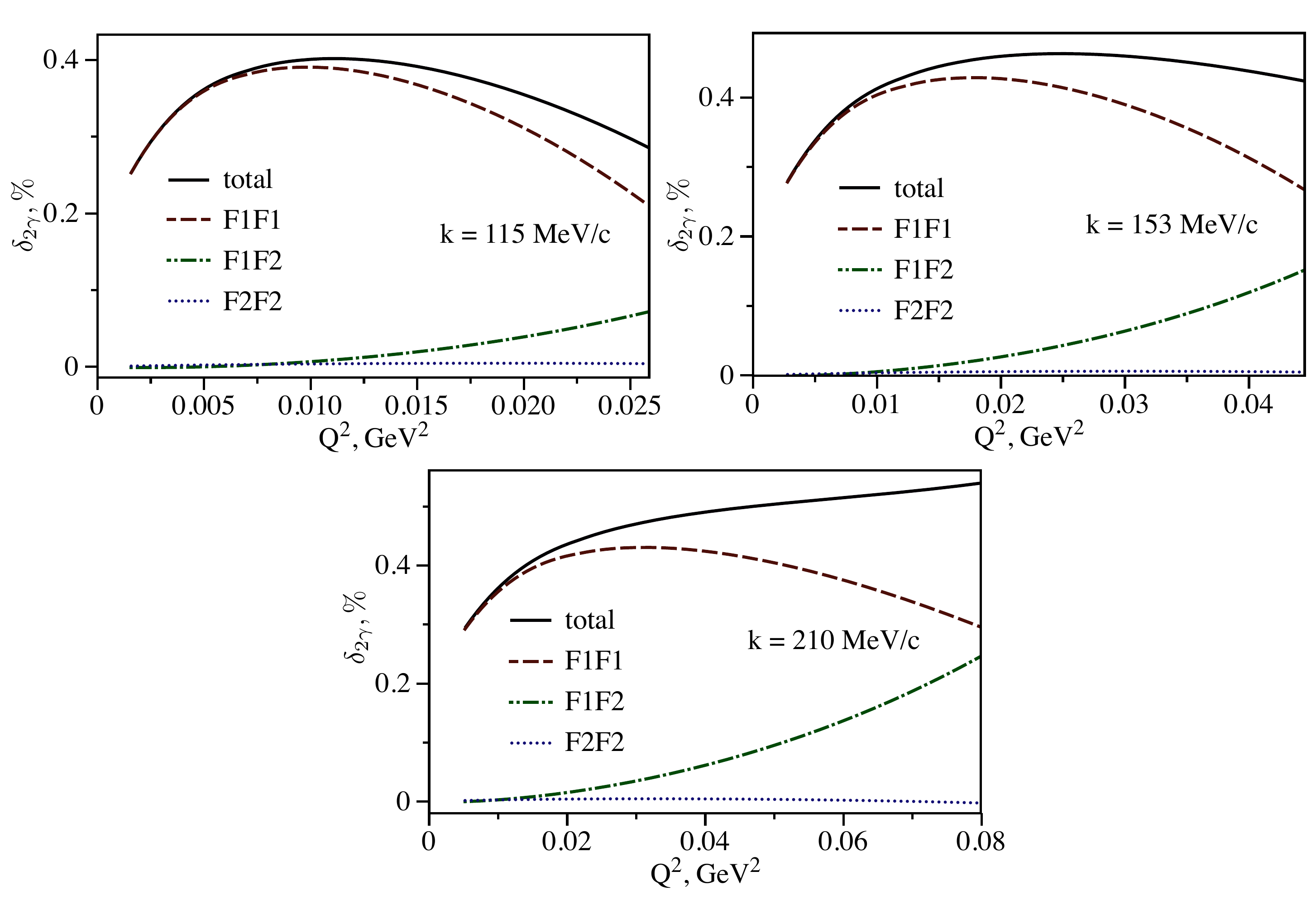}
\end{center}
\caption{ TPE correction to the unpolarized elastic $ \mu^{-}p $ cross section for three different muon beam momenta. The total correction is shown by the black solid curves, the contribution from the F1F1 structure of photon-proton-proton vertices is shown by the red dashed curves, the contribution from the F1F2 structure by the green dashed-dotted curves, and the contribution from the F2F2 structure by the blue dotted curves.}
\label{tpe_FFs}
\end{figure}

The predictions of the TPE corrections in the elastic muon-proton scattering in terms of the different vertex structures are shown on Fig. \ref{tpe_FFs} for the MUSE experiment kinematical region.

One notices from Fig. \ref{tpe_FFs} that the F2F2 vertex structure does not contribute significantly to the cross section, while the main contribution comes from the F1F1 vertex structure. The contribution from the F1F2 vertex structure rises when increasing the momentum transfer. This contribution is significant only for largest values of momentum transfer of the MUSE experiment. In magnitude, the TPE correction varies between $ 0.25 \% $ and $ 0.5 \% $.

\begin{figure}[htp]
\begin{center}
\includegraphics[width=1.\textwidth]{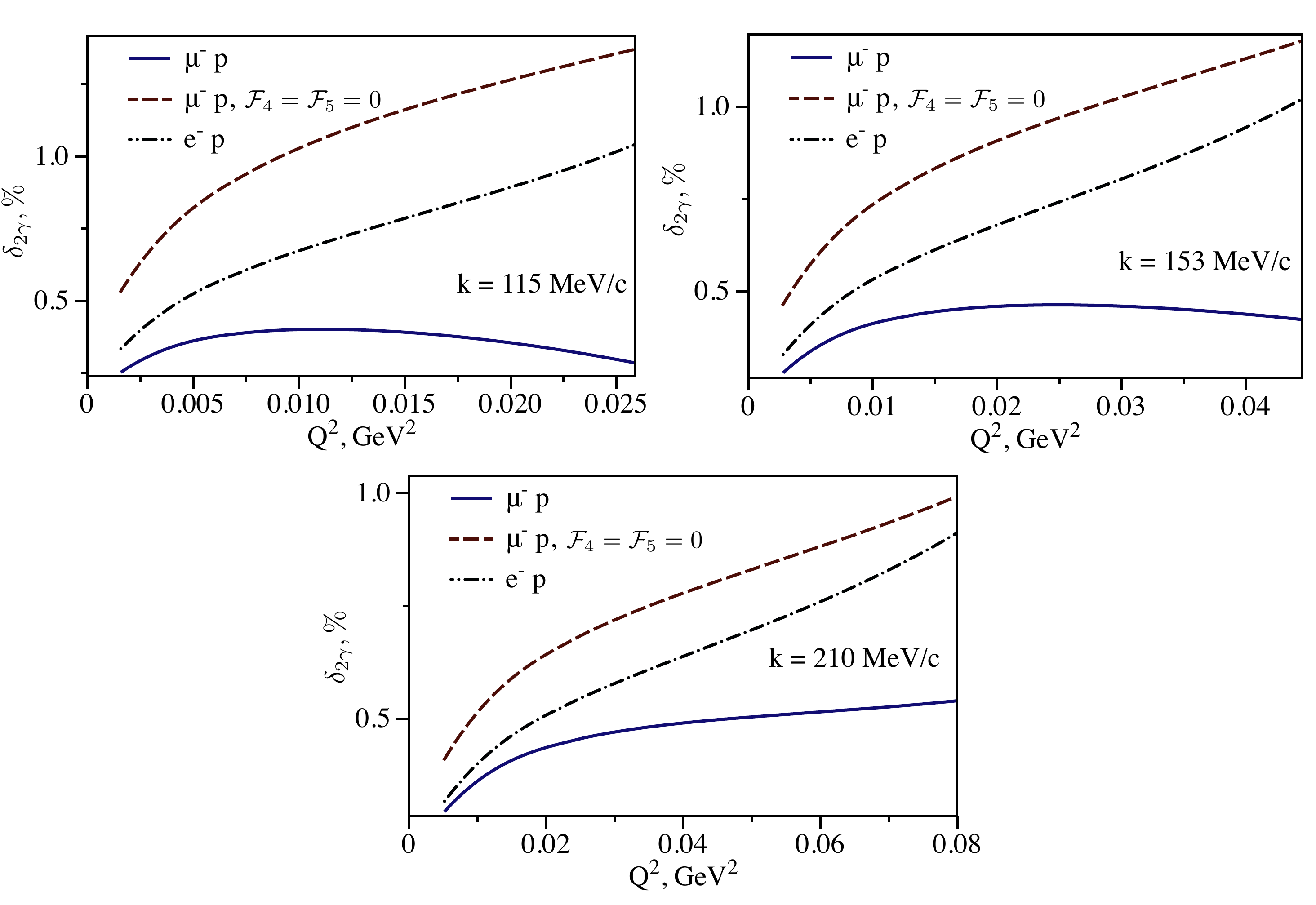}
\end{center}
\caption{TPE correction to the unpolarized cross section for three different muon beam momenta. The TPE correction to elastic $ \mu^{-} p $ scattering is shown by the blue solid curves, the black dashed-dotted curves show the elastic $ e^{-} p $ scattering correction, the elastic $ \mu^{-} p $ scattering correction without account of muon helicity flip is shown by the red dashed curves.}
\label{tpe_electron_muon}
\end{figure}

We show a comparison between the TPE corrections to elastic electron-proton and elastic muon-proton scattering in Fig. \ref{tpe_electron_muon}. One sees that the TPE correction in the case of muon-proton scattering is smaller than the correction in the case of electron-proton scattering with the same lepton beam momenta. The contribution of the helicity-flip amplitudes plays a significant role for $ \mu^{-}p $ scattering in the kinematical region of the proposed experiment. It contributes with a sign opposite from the contribution of the amplitudes without helicity flip and significantly reduces the correction. We found that for the higher momentum transfer $ Q^2 \sim 1-2 ~\mathrm{GeV}^2 $ the contribution of helicity flip amplitudes does not play a significant role and the predictions for $ \mu^{-} p $ elastic scattering only slightly deviate from the predictions for $ e^{-} p $ elastic scattering, in agreement with the findings of Ref. \cite{Chen:2013udl}.

\begin{figure}[htp]
\begin{center}
\includegraphics[width=.70\textwidth]{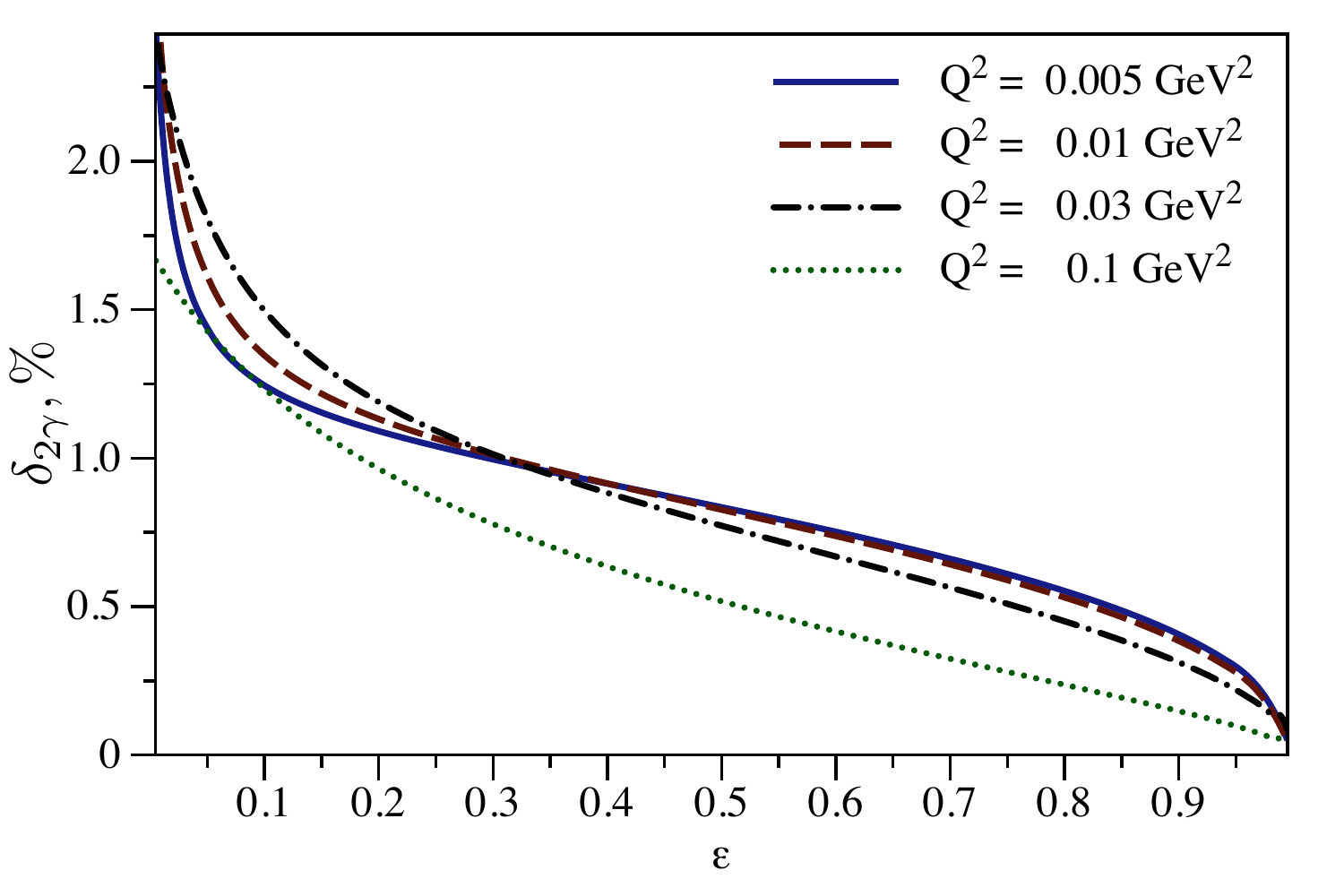}
\end{center}
\caption{$ \varepsilon $ dependence of the TPE correction to the unpolarized cross section for the elastic electron-proton scattering for different momentum transfers.}
\label{del}
\end{figure}

\begin{figure}[htp]
\begin{center}
\includegraphics[width=0.95\textwidth]{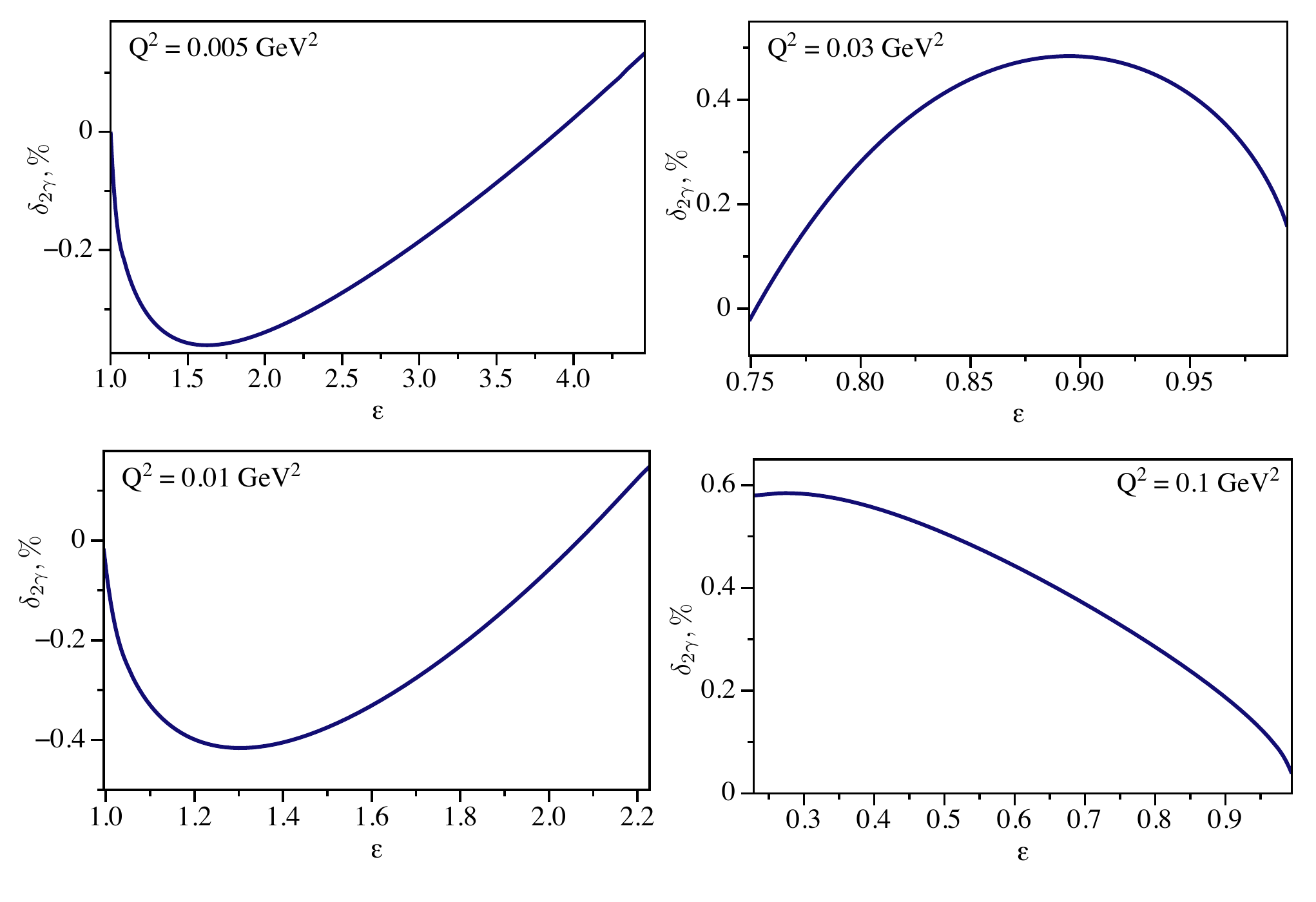}
\end{center}
\caption{$ \varepsilon $ dependence of the TPE correction to the unpolarized elastic $ \mu^{-}p $ cross section.}
\label{muon_fixed_q2}
\end{figure}

We show the TPE corrections as a function of $ \varepsilon $ for the fixed momentum transfer in Figs. \ref{del} and \ref{muon_fixed_q2}. The results for elastic electron-proton scattering in the zero electron mass limit are nearly indistinguishable from the results with finite electron mass, which are shown in Fig. \ref{del}. These results are in agreement with the results of previous calculations \cite{Bernauer:2010wm}, which were based on Ref. \cite{Blunden:2003sp}. The slight difference between our results comes from the different parametrizations of electric and magnetic form factors in our work and in Ref. \cite{Blunden:2003sp}.

\section{Conclusions and outlook}
\label{sec5}

In this work we have extended the general formalism of TPE corrections to the elastic unpolarized scattering of a finite mass lepton off a nucleon target. We have estimated the cross section correction for the future MUSE experiment in a model for the TPE correction with proton intermediate states. The estimates for the TPE correction of the muon-proton scattering cross section vary between $ 0.25 \% $ and $ 0.5 \% $. These estimates are up to a factor three smaller, as compared with TPE corrections for the case of electron-proton elastic scattering in the same lepton kinematical region. This is due to the contribution of lepton helicity-flip amplitudes, which have an opposite sign as compared with the contribution of nonflip amplitudes and significantly reduce the correction. To go to larger momentum transfer, a next step will be to include inelastic state contributions within a dispersive formalism.

\appendix

\section{The relation between helicity amplitudes and structure amplitudes}
\label{app1}

Using the Jacob and Wick \cite{Jacob:1959at} phase convention for the spinors, the helicity amplitudes $ T_{h' \lambda', h \lambda} $ for elastic lepton-nucleon scattering are expressed in terms of the generalized FFs by
\ber \label{hamp}
 \frac{\Sigma \xi^2}{e^2} T_1 & = & 2 ( \frac{\Sigma Q^2}{ \Sigma - s Q^2} + s - M^2 - m^2 ) \cG_M - 2 (s-M^2-m^2) \cF_2 +  \frac{(s-M^2-m^2)^2}{M^2 } \cF_3 +  \nonumber \\
& & 4 m^2 \cF_4 + 2 m^2 \frac{s-M^2-m^2}{M^2} \cF_5, \nonumber \\
 \frac{M \Sigma}{e^2} \xi T_2 & = &   2 M^2 (s - M^2 + m^2)  \cG_M - ( (s - m^2)^2 - M^4) \cF_2 + ( (s - M^2)^2 - m^4 ) \cF_3  + \nonumber \\
 & &   2 (s + M^2-m^2) m^2 \cF_4 + 2  (s - M^2 + m^2) m^2 \cF_5,  \nonumber \\
  \frac{\Sigma \xi^2}{ e^2} T_3& = & 2 (s - M^2 - m^2)   ( \cG_M - \cF_2 ) +  \frac{ (s - M^2 - m^2)^2 }{ M^2} \cF_3 + 4 m^2  \cF_4 +  2 \frac{m^2 (s - M^2 - m^2) }{ M^2 } \cF_5,  \nonumber \\
  \frac{\Sigma }{m e^2} \xi T_4& = &  - 2 (s + M^2 - m^2)  (  \cG_M - \cF_2 )  - \frac{  ( (s - m^2)^2 - M^4)  }{ M^2} \cF_3 - 2 (s - M^2 + m^2)  \cF_4 - \nonumber \\
& &   \frac{  ( (s - M^2)^2 - m^4 ) }{ M^2} \cF_5,  \nonumber \\
 \frac{M \Sigma}{m e^2}  T_5& = &   - 4 M^2 s \cG_M +   (s + M^2 - m^2)^2 \cF_2 -   ( s^2 - (m^2 - M^2)^2 ) ( \cF_3  + \cF_4)  -  \Sigma \cF_6-\nonumber \\
& &  ( s - M^2 + m^2 )^2 \cF_5,	  \nonumber \\
  \frac{M \Sigma}{m e^2} T_6& = &   4 M^2 s \cG_M -  (s + M^2 - m^2)^2 \cF_2 + ( s^2 - (m^2 - M^2)^2 )  ( \cF_3   + \cF_4 ) - \Sigma \cF_6   + \nonumber \\
& &  ( s - M^2 + m^2 )^2 \cF_5.	 
\eer

These relations can be inverted to yield the generalized FFs in terms of the helicity amplitudes $ \tilde{t} = \frac{T}{e^2} $  as
\ber \label{stramp}
 \cG_M & = & \frac{1}{2} ( \tilde{t}_1 - \tilde{t}_3 ), \nonumber \\
 \Sigma \cF_2 & = & - 2 m^2 M^2  \tilde{t}_1  - M (\left(s-M^2\right)^2-m^4) \xi \tilde{t}_2  - M^2 \eta(m) \tilde{t}_3 + 2 m M^2  \left(s -M^2+m^2\right) \xi \tilde{t}_4 - \nonumber \\
& &   m M 
   \left(s - m^2 - M^2\right) ( \tilde{t}_5 - \tilde{t}_6), \nonumber \\
 \frac{ \Sigma }{M^2} \cF_3 & = & - (s-m^2-M^2) \tilde{t}_1 - 2
   M \left(s -M^2+m^2\right) \xi  \tilde{t}_2 + \rho_3 \tilde{t}_3  + 2 m \left(s+M^2-m^2\right) \xi  \tilde{t}_4  -  \nonumber \\
& &  2 m M( \tilde{t}_5 - \tilde{t}_6),   \nonumber \\
 \frac{ \Sigma }{ M } \cF_4 & = & - M \left(s-m^2-M^2\right) \tilde{t}_1 - (\left(s -m^2\right)^2-M^4) \xi  \tilde{t}_2  + M \rho_3 \tilde{t}_3 +  \frac{M (\left(s -M^2\right)^2-m^4)}{ m } \xi  \tilde{t}_4 - \nonumber \\
& &     \frac{  \left(s-m^2-M^2\right)^2}{2 m} ( \tilde{t}_5 - \tilde{t}_6),  \nonumber \\
 \frac{\Sigma}{M^2} \cF_5 & = & 2 M^2 \tilde{t}_1 + 2 M \left(s+M^2-m^2\right)\xi  \tilde{t}_2 +  \eta(M) \tilde{t}_3 - \frac{\left(s-m^2\right)^2-M^4}{ m} \xi  \tilde{t}_4 + \nonumber \\
& &   \frac{M \left(s- m^2 -M^2\right)}{m}( \tilde{t}_5 - \tilde{t}_6),  \nonumber \\
 \cF_6 & = & - \frac{M}{2m} ( \tilde{t}_5 + \tilde{t}_6 ),
\eer
with
\ber
 \xi & = & \sqrt{\frac{Q^2}{\Sigma - s Q^2}}, \nonumber \\
\eta(m) & = &  \frac{ 2 m^2 \left(\Sigma + s Q^2 \right)+\Sigma Q^2}{s Q^2-\Sigma }, \nonumber \\
 \rho_3 & = & \frac{ \Sigma m^2 + \left(m^2-M^2\right)^2 \left(M^2+Q^2\right)+s^2 \left(2 m^2+3 M^2\right)-s \left(m^4+Q^2 \left(m^2+M^2\right)+3 M^4\right)-s^3}{  s Q^2 -\Sigma }. \nonumber
 \eer

\section*{Acknowledgements}

This work was supported in part by the Deutsche Forschungsgemeinschaft DFG in part through the Collaborative Research Center [The Low-Energy Frontier of the Standard Model (SFB 1044)], in part through the Graduate School [Symmetry Breaking in Fundamental Interactions (DFG/GRK 1581)], and in part through the Cluster of Excellence [Precision Physics, Fundamental Interactions and Structure of Matter (PRISMA)].

\end{document}